\documentclass[oldversion]{aa}
\usepackage{graphicx}
\usepackage{natbib}
\usepackage{txfonts}
\begin{document}
    \title{Medium resolution 2.3$\mu$m spectroscopy of the massive  
Galactic open cluster Westerlund~1\thanks{Based on observations
	  collected at the European Southern Observatory, Chile, under
	  program number 076.D-0680(A)}}

\titlerunning{NIR spectroscopy of Westerlund 1}

    \author{S. Mengel
           \inst{1}
           \and
           L.E. Tacconi-Garman\inst{1}
           }

    \offprints{S. Mengel}

    \institute{European Southern Observatory (ESO),
               Karl-Schwarzschild-Str. 2,
	      D-85748 Garching\\
               \email{smengel@eso.org}
              }

    \date{Received ; accepted }


\abstract
    {The Galactic open cluster Westerlund~1 was found only a few  
years ago to be much more massive than previously thought, with evidence 
suggesting its mass to be in excess of $\sim$10$^5$\,M$_\odot$\@, 
in the range spanned by young extragalactic star clusters. 
Unlike those clusters its proximity makes spatially resolved studies of its  
stellar population feasible. It is therefore the ultimate template for a young, 
massive star cluster, permitting direct comparison of its properties with 
measurements of velocity dispersion and dynamical mass for spatially unresolved 
extragalactic clusters.

    To this end, we used the long slit near-infrared spectrograph  
VLT/ISAAC to observe the CO bandhead region near 2.29$\,\mu$m,
scanning the slit across the cluster centre during the integration.
Spatially collapsing the spectra along the slit results in a  
single co-added spectrum of the cluster, comparable to what one would obtain in  
the extragalactic cluster context.
     This spectrum was analysed the same way as the spectra of almost point-like 
extragalactic clusters, using red superiant cluster members as velocity templates.
     We detected four red supergiants which are included in the integrated 
spectrum, and our measured  velocity dispersion is 5.8\,km/s. Together with the 
cluster size of 0.86\,pc, derived from  archival near-infrared SOFI-NTT
images, this yields a dynamical mass of 6.3$\times10^4$M$_\odot$. While this 
value is not to be  considered the final word, there is at least so far no sign 
for rapid expansion or collapse.

    \keywords{open clusters and associations: individual: Westerlund 1 -- Galaxies: star clusters -- supergiants}
}
    \maketitle
%

\section{Introduction}
For more than a decade now, following their discovery by
\cite{Hetal92}, young massive star clusters have been the
focus of intense studies, amongst other reasons because they  
represent excellent
laboratories of star formation: their relative youth ensures that their
present day mass function still closely resembles the initial mass  
function (IMF),
their high mass in principle allows studies of IMF variations  
(stochastic sampling of the IMF
plays a minor role in clusters with masses in excess of approximately  
10$^5$M$_\odot$), and last, but not least, they are abundant in
different types of galaxies where they probe a variety of environments.
They suffer only one major drawback, the large distance, which means  
that
they are slightly resolved at most, and typically also fairly faint.

The ideal candidates for star formation studies would show the  
advantages of
the extragalactic clusters, but sit in the Galaxy and consequently allow
spatially resolved analysis of individual stars in the clusters.
Westerlund~1 (Wd~1) is potentially such an ideal target, the first  
one discovered in the
Milky Way (at a distance of 5 kpc, where 1\arcsec\ corresponds to 0.025 pc). 
Even though it was already detected
in the 1960's \citep{Westerlund61}, it took a few decades to find out  
that the cluster light is highly extinguished by large quantities of  
dust, and that
it possibly has a fairly high mass of about 10$^5$M$_\odot$ .

We observed it as a template for the almost unresolved extragalactic  
star
clusters. For most of those which were analysed spectroscopically,
it is plausible that a large fraction of the cluster mass is contributed
by low-mass stars (e.g., \citet{Mengel02, LarsenRichtler04,  Larsenetal04}),
and that the clusters are compact enough to survive for at least several
Gyrs. On the other hand, age determinations of individual clusters in
a cluster population suggest that around 90\% of all young massive
extragalactic star clusters, irrespective of mass, are disrupted 
within 10 Myrs \citep{Fall04, Mengel05, Bastian05, Whitmore06}.
Of course, it is possible that the clusters targetted for spectroscopy
have already undergone a selection effect, because they are all older than
6-7 Myr and therefore clearly not amongst the first ones to disrupt. Other types
of investigation are necessary to test this hypothesis.

But the origin of this discrepancy could also lie in false
assumptions concerning the dynamical state of the young extragalactic
star clusters - for example, disruption or collapse could lead to
an increase in velocity dispersion which is unrelated to cluster mass.
Therefore we wanted to test the approach used for extragalactic clusters on the  
only spatially resolved young
star cluster which has a mass around $10^5$M$_\odot$.


\section{Observations and data reduction}

\begin{figure}
    \centering
     \includegraphics[bb=216 190 479 554,width=8cm,angle=0,clip]{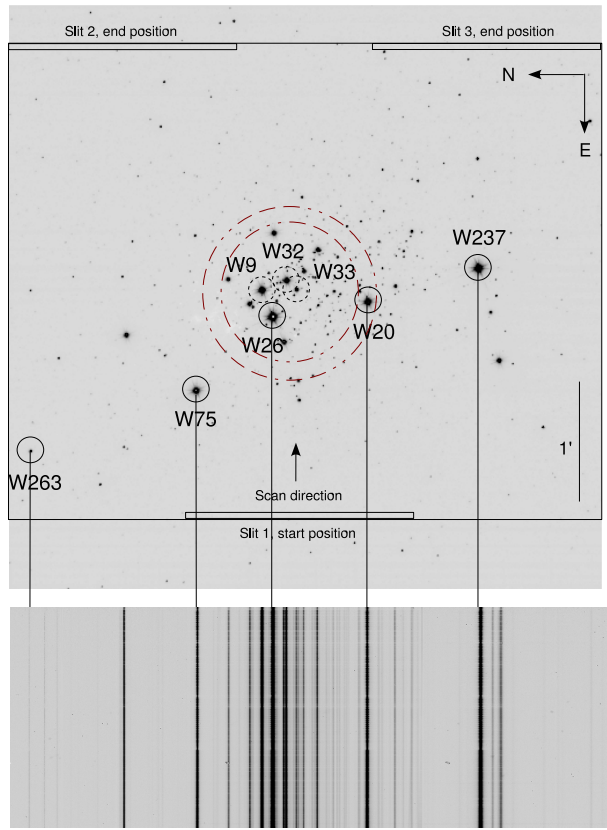}
       \caption{Top: 2MASS K-band image of the centre of Westerlund~1.
	       Indicated is the region over which the slit was scanned.
	       Three scans were performed, each starting on the eastern
		 edge and scanning the slit across the field of view during
		 the five minute integration. 
	       The brightest stars are saturated in the image. Small black full circles
	       are late-type stars, small black broken circles are the NIR-brightest
	       early type stars, and the large broken red (grey) circles
	       indicate the half-light radius (including uncertainties).
	       Bottom: Mosaic of the three spatial scans of Westerlund 1, 
	       integrated over 4$^\prime$ perpendicular
	       to the slit. Increasing
	       wavelength from bottom to top. Sky subtracted, flatfielded and
	       distortion corrected, aligned with image above. This 2d frame
	       was collapsed along the x-axis in order to obtain the 1d spectrum.
                    \label{scanimage}     }
    \end{figure}

Observations were conducted on VLT-ANTU on the nights of March 11 and  
12, 2006
and consisted of scanning the slit across a region while integrating.
The ISAAC slit has a length of 2$^\prime$, and we selected a slit  
width of 0\farcs3
to obtain the highest possible spectral resolution. This is necessary  
for determining a stellar velocity dispersion below 10\,km/s, which we  
expected for the cluster.
The scans were conducted by acquiring a star in the centre of the cluster
(W33 - throughout the text, we use the numbering from \citet{Westerlund87}),
offsetting 2$^\prime$ East, and then scanning westward over 4\arcmin,
for a total integration time of 300\,s. This was repeated three times,
and two more scans were conducted by offsetting 108$^{\prime\prime}$  
North and South, respectively,
from the acquisition position.
Figure \ref{scanimage} shows the region covered and the resulting 
2-dimensional spectrum.

The central wavelength was set to be 2.31\,$\mu$m, and the actual  
wavelength range
covered is 2.249 -- 2.373\,$\mu$m at a resolution of R$\approx$9000.

Reduction of the data was performed using the IRAF
package\footnote{IRAF is distributed by the National Optical Astronomy
Observatories, which are operated by the Association of Universities
for Research in Astronomy, Inc., under cooperative agreement with the
National Science Foundation.}.
For the slit scan, it included sky subtraction (a scan across an  
empty sky
field 10$^\prime$ away was performed with identical integration time  
of 300\,s directly
after the cluster scan), bad pixel correction (but there was a blotch of
bad pixels (dust) in the top left quadrant which appeared during the night  
which neither
flatfielded out, nor was properly removed by median interpolation),  
and flatfielding.

Wavelength calibration combined the information of the XeAr lamp  
spectra and the night sky lines \citep{Rousselotetal}. 
Spatial distortion was easily corrected due to the  many star spectra
along the slit. The three scans (centre, offset North and offset  
South) were mosaicked into a combined scan before integrating spatially 
into a  single cluster spectum.

Individual spectra of the four near-infrared brightest stars  
(indicated with small solid circles in Figure \ref{scanimage}), which show prominent
CO absorption, were extracted as template stars for the determination  
of the cluster velocity dispersion.

Since solar type stars were observed as telluric standards (Hip077718  
(G2V, m$_K$=5.894) and Hip087370 (G3V, m$_K$=5.609)),
we divided them by an adequately smoothed high-resolution solar  
spectrum before applying the correction of telluric features to the Westerlund~1  
spectrum.
The features near the MgI line around 2.2814\,$\mu$m appear to be a  
residual from this step which we could not eliminate.

Finally, we normalized all spectra which were used for the analysis  
by dividing by the continuum level determined from a small window (roughly 50  
pixels) just shortward of the CO bandhead.
A rough flux calibration was obtained via a
two-step approach using the archival SOFI/NTT narrow band data  
obtained for \citet{Crowtheretal2006}, programme ID 073.D-0321, and the relative  
intensities from our scan. Even though we used a combination of NB 2.09$\mu$m and  
NB 2.17$\mu$m,
we could reproduce the magnitudes published by \citet{Crowtheretal2006} to within
0.1mag, and determined approximate K-band magnitudes for the  moderately bright
cluster members not listed in their publication. For the brightest cluster  
members, we could
not use this image, because they were saturated. In those cases we  
used relative intensities
to fainter stars common to the scan and the image (and isolated  
enough such that
spectra of different clusters were not projected on top of each other).
The relative fluxes of the stars remained quite  
constant throughout the short scanning time of 5 minutes, because the
seeing varied only slightly. The three individual scans of the central region
showed variations in the relative stellar intensities of up to 20\%, more
typically around 10\%.

\section{Analysis and Results}\label{analysisresults}

\begin{figure}
\centering
\includegraphics[bb=210 237 496 665, width=8cm,angle=0, clip]{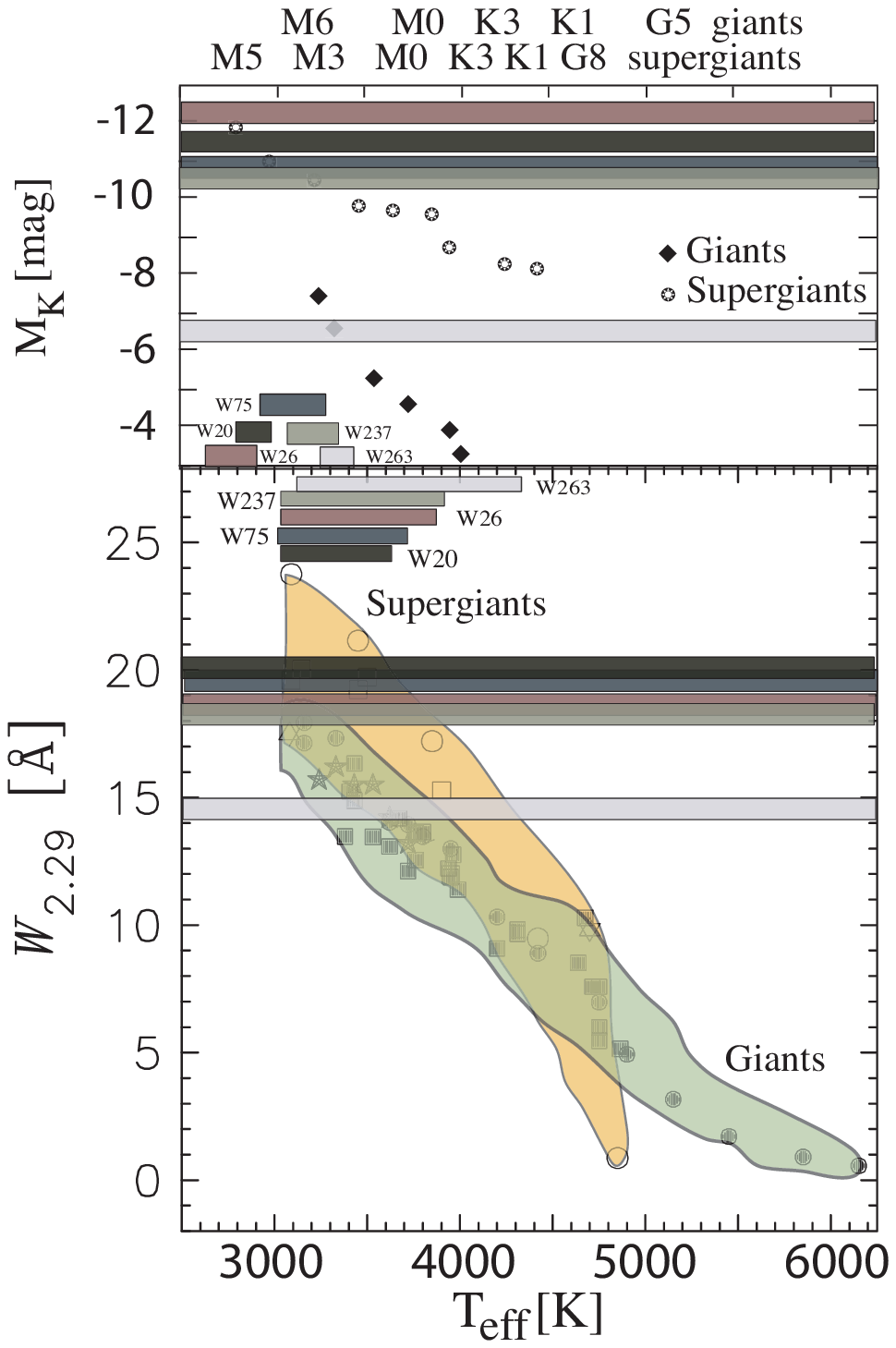}
    \caption{Determination of the spectral types of the observed stars. 
Our measured values for  W$_{\mathrm{CO}2.29}$ and M$_K$ are compared to published 
measurements as a function of T$_{eff}$ \citep{Foerster2000, Koornneef83, SchmidtKaler82}. 
For details, see text.
  \label{sptype}}
\end{figure}

\begin{figure*}
\centering
\includegraphics[bb=27 37 591 755, width=9cm,height=17cm,angle=-90,clip]{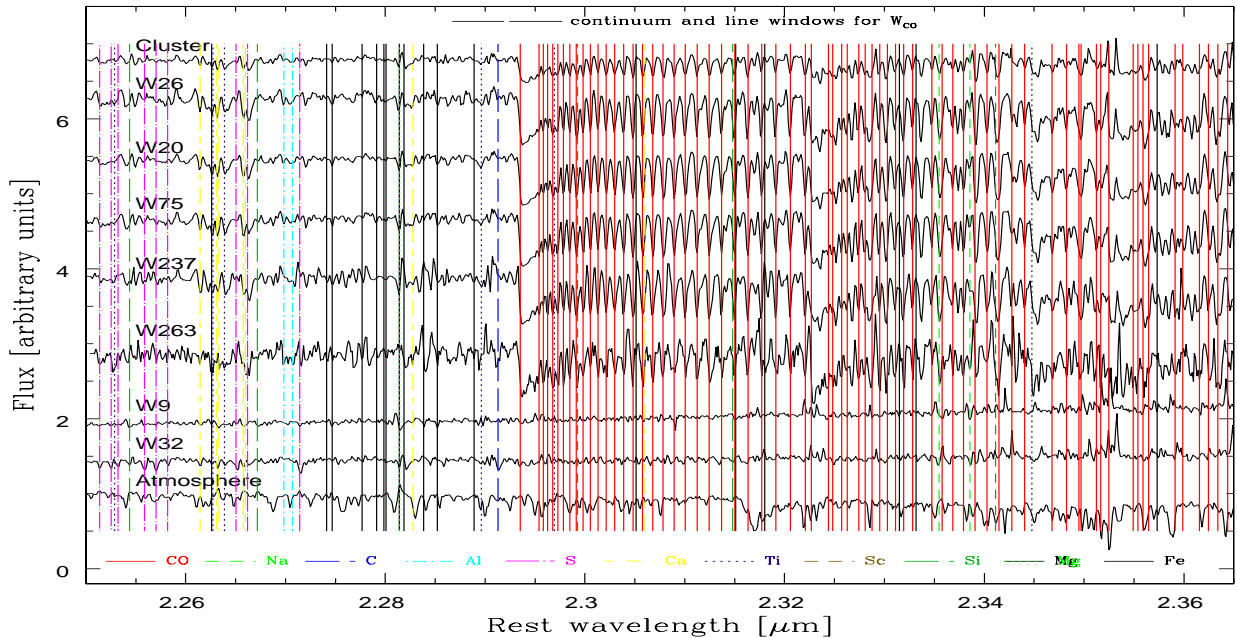}
    \caption{Specta of integrated cluster and the near-infrared
	       brigthest individual stars. Atmospheric absorption
             (corrected for, but over-compensation and sub-optimal sky subtraction
              causes a few artefacts at long wavelengths) is shown at  
the bottom. Line identifications taken from \citet{Hinkle95}.
       \label{combinedspectra}}
\end{figure*}

The resulting spectra are shown in Fig.~\ref{combinedspectra}. We  
additionally indicate
the region which was used to determine the equivalent width of the  
2.29$\mu$m CO bandhead
according to the definition of \cite{OO00}.
From this equivalent width W$_{\mathrm{CO}2.29}$, in combination with
approximate absolute magnitudes,  we determined the rough spectral  
type of the
individual stars as explained in Fig. \ref{sptype}.

For the K-band magnitudes, we used a distance modulus of
13.5,  m$_V$ from \citet{Westerlund87}, and apply an extinction correction
of 0.7 mag in K-band, which is a lower limit \citep{Crowtheretal2006}.

Fig. \ref{sptype} shows our measurements of W$_{\mathrm{CO}2.29}$ in comparison
with published values as a function of T$_{eff}$. The lower
part of the figure is based on a figure from \citet{Foerster2000} which includes
measured values for W$_{\mathrm{CO}2.29}$  for giants and RSGs from her own measurements 
and two other publications (see references in that work). The loci populated by giants and
SGs are indicated as shaded areas, and our measured values for both, W$_{\mathrm{CO}2.29}$  and
M$_K$ are represented as horizontal bars which also indicate the measurement uncertainties.
The range of effective temperature for which a given measurement overlaps with measured
values for stars is indicated by the shorter (labelled) horizontal bars in the top of the
lower part of the figure. In the top part, we applied essentially the same technique.
For theoretical M$_K$, we used V$-$K colours from  \citet{Koornneef83},
and theoretical M$_V$ from \citet{SchmidtKaler82}). For three stars, both diagnostics
lead to an agreement in spectral type / T$_{eff}$. For the two brightest stars, the mismatch
is caused by the lack of data points in \citet{Foerster2000} below 3000 K. Therefore,
the table lists the spectral type resulting from M$_K$.    

The four brightest stars are assigned spectral types ranging between M3I and M6I (foreground
objects are ruled out also because of the small spread in radial velocity).
In order to determine if W263 is a  
member of Wd~1, and hence valid for inclusion in our velocity  
dispersion estimate, we determined
its spectral type and radial velocity. It is much fainter than the  
red supergiants, and we considered
W$_{\mathrm{CO}2.29}$ and
V- and K-band photometry to decide if it is a cluster member or not.
The equivalent width would be consistent with a late K-type  
supergiant or a late M giant.
But the absolute luminosity is much too faint, an M5III star  
approximately at the distance
of Wd~1 is likely, or a similar spectral type if it is somewhat  
closer or more distant.
Moreover, its radial velocity is very different from that of the  
other stars; the velocity
difference of around 60\,km/s would be several times the escape  
velocity of the cluster.  Hence, we have rejected W263 from our  
dynamical analysis.

We compared W$_{\mathrm{CO}2.29}$ = 8.7\AA\ of the integrated spectrum with predictions from
Starburst99 \citep{VazquezLeitherer05}, using an instantaneous burst at solar
metallicity with a Kroupa  IMF ($\alpha=1.3$ between 0.1 and 0.5 M$_\odot$,
$\alpha=2.3$ between 0.5 and 100 M$_\odot$, \citet{Kroupa2002}). The modelled W$_{\mathrm{CO}2.29}$ 
reaches 8.7\AA\ at an age of 6.3 Myr. It is difficult to quote uncertainties for this
age, because the main uncertainty is not caused by measurement uncertainties in the
equivalent width, but rather in the models (e.g. treatment of the post-main-sequence phase),
the assumptions (instantaneous burst), or simply statistics. Therefore, while the
uncertainty in the equivalent width would lead to an age uncertainty of only around 0.2 Myr,
the real age may easily be off by one or two Myr.
Some of the spectral properties of the cluster as a whole and of  
individually extracted stars are listed in Table~1.

\begin{table}
\begin{center}
\begin{tabular}{lcccc}
Spectrum	&	W$_{\mathrm{CO}2.29}$	&	 m$_K$     & M$_{\mathrm{K}_0}$ &  Spectral Type / age\\\hline
                 &       \AA             &        mag    & mag    &                \\\hline\hline
Westerlund 1 	&	8.7		&        0.1       &     -     &    6.3 Myrs	\\
W26	        &	18.6		&        1.9       &   -12.3   &    M5-6I         \\
W20		&	20.3		&	 2.7       &   -11.5   &    M5I	\\
W75		&	19.5		&	 3.3       &   -10.9   &    M4I         \\
W237            &       18.2            &        3.4       &    -10.6   &    M3I         \\\hline
W263            &       14.5            &        7.2       &    -6.5    &    M5III       \\\hline
\end{tabular}\label{spectratable}
\caption{Properties of the integrated cluster spectrum and of the  
five late-type stars
included in that spectrum. The columns list: Designation as in \cite{Westerlund87},
W$_{\mathrm{CO}2.29}$ according to the definition in \cite{OO00}, the  
apparent and absolute
K-band magnitudes (where the uncertainty is around 0.3 mag for the  
first, and 0.4 mag for
the latter) and the spectral type of the
stars determined from W$_{\mathrm{CO}2.29}$ and the absolute  
magnitude. For the cluster,  
the last column lists the cluster
age determined from  W$_{\mathrm{CO}2.29}$  in comparison to  
predictions from Starburst99 \citep{VazquezLeitherer05}.
The apparent magnitude listed for the integrated spectrum is only  
given in order
to be able to compute the relative contributions of the individual  
stars to the total
spectrum from it. W26, the NIR brightest object, contributes 20\%,  
and all the late type
stars together around 40\%. The four brightest stars have magnitudes
identifying them as red supergiants at the distance of Wd 1.}
\end{center}
\end{table}

The integrated spectrum was used together with the individual spectra  
to determine the
velocity dispersion in the same manner as with spectra of  
extragalactic star clusters: broadening
the continuum-diluted red- or blueshifted template spectra\footnote{Velocity dispersions
from broadening of absorption features can only be measured for clusters
dominiated by a stellar type with photospheric absorption and without fast rotation} by  
convolution with a Gaussian with a given
velocity width
until a best fit (determined through a $\chi^2$ minimization) of the  
cluster spectrum is obtained.
 From the estimated total mass of the cluster (10$^5$M$_\odot$, 
\citet{Clarketal2005}), we expected a velocity dispersion of around 9\,km/s, still
quite comfortably above the limit of around 7\,km/s to which we can  
determine a velocity dispersion
in that spectral range if the signal-to-noise ratio is high enough  
(see Fig. 4, top, in \citet{Mengel02}).

However, our measured velocity dispersion is much lower than expected.
For different template stars (which were of course the individual  
stars extracted from
the scan) and different wavelength ranges, our fitting routine
returned results  
between 1 and 6\,km/s.
As there was no trend in our determined value with either spectral  
type of the template or spectral region considered, 
it was obvious that this was the random scatter expected
for a velocity dispersion below the resolution limit, and we can only  
deduce that the velocity dispersion is below 7\,km/s.

\begin{figure}
\centering
\includegraphics[bb=27 37 591 755, width=5.5cm,angle=-90, clip]{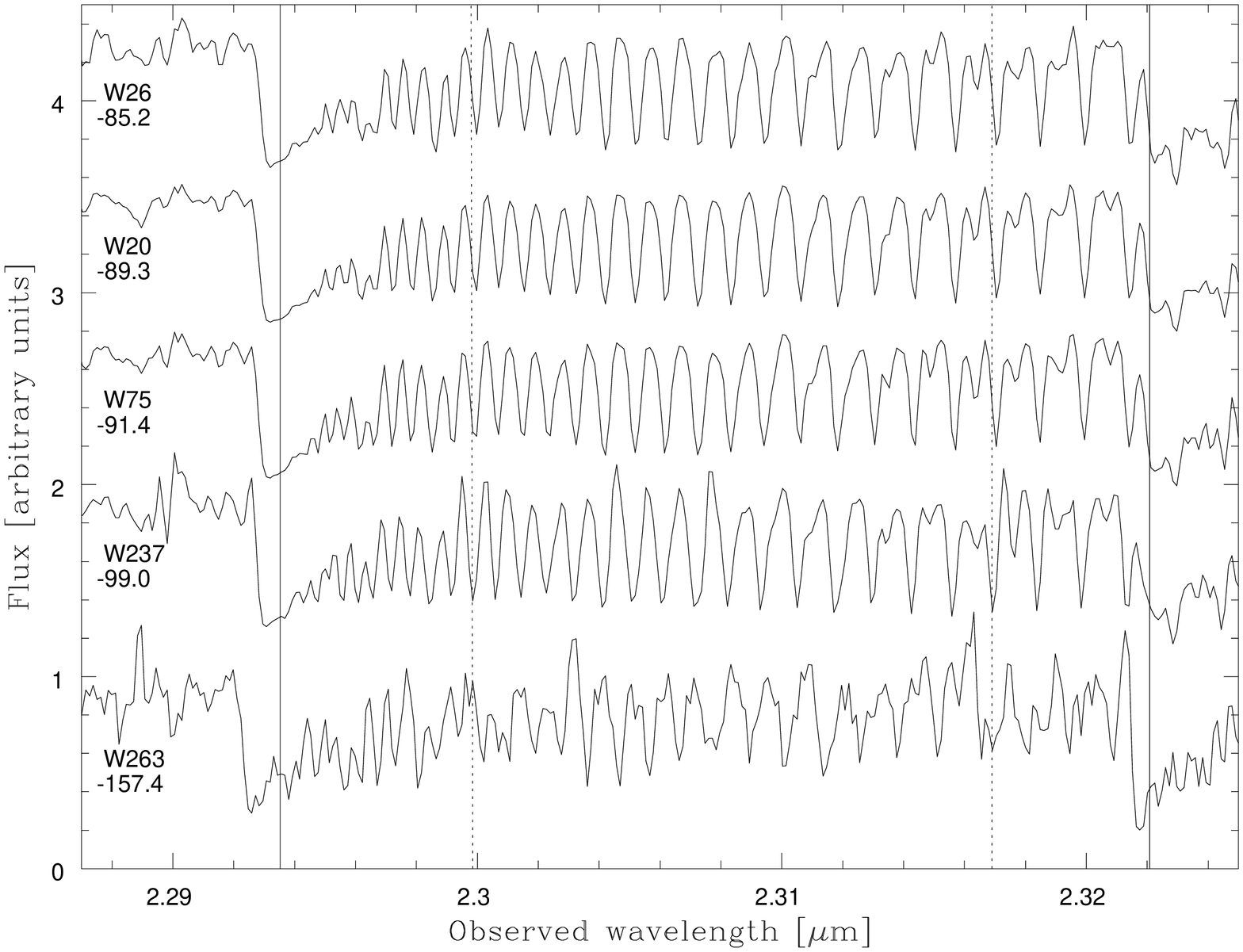}
    \caption{Spectra of  the   brightest red supergiants and the red  
giant not corrected for radial velocity. The lines (bold: CO 2-0 bandhead and 
CO(2-0) R13 rest  wavelengths, dashed: centerd on two arbitrary absorption lines 
in the lowest red  supergiant spectrum) are supposed to guide the eye, showing that 
a relative velocity between the  stars can easily be determined. Below the names, 
heliocentrically  corrected radial velocities are given.
       \label{obsspectra}}
\end{figure}

What is possible to determine, though with relatively high  
statistically uncertainty, is the rms radial velocity dispersion of  
the four
red supergiants.
The rms velocity dispersion determined from the - unfortunately only  
four - individual
stellar velocities is 5.8$\pm$2.1\,km/s. Weighting by the luminosity  
lowers this value to
5.0$\pm$1.7\,km/s.  Errors on both of these values were determined  
through Monte Carlo simulations, using the uncertainties in the  
individual velocity determinations.
Formally, the latter value would be the one to compare with values  
obtained from observations of extragalactic clusters.  But since this  
comparison is inappropriate in this case, due to the fact that
Westerlund~1, because of its young age, contains far fewer red 
supergiant (RSG) stars than all extragalactic  
clusters studied so far, we use the
dynamically more meaningful unweighted rms.
Apparent magnitudes of the stars are given in Tab.~1, and the
heliocentrically corrected radial velocities in Fig. \ref{obsspectra}.

The other parameter we require for the determination of the dynamical  
mass is the
projected half-light radius.  We used the NIR image we created from  
the archival NTT/SOFI
data to obtain a radius which contains half of the total light (in  
projection) to be
r$_{hp}$=0\farcs86$\pm$0\farcs14. The uncertainty comes mostly  
through assumption of
different locations of the weakly constrained cluster centre.

The resulting dynamical mass, for a cluster in Virial 
equilibrium,  is (assuming $\eta = 9.3$, for details  
on the formula see
\citet{Mengel02})
\begin{displaymath}
M_{dyn} = \eta \sigma^2 r_{hp} / G = 2161\times 5.8^2 \times 0.86 M_\odot = 6.3^{\scriptstyle{+5.3}}_{\scriptstyle{-3.7}}\times 10^4\,M_{\odot}
\end{displaymath}
Within the errors this is consistent with the mass expected from  
photometry \citep{Clarketal2005}.

\begin{figure}
\centering
\includegraphics[bb=27 37 591 755, width=6cm,angle=-90, clip] 
{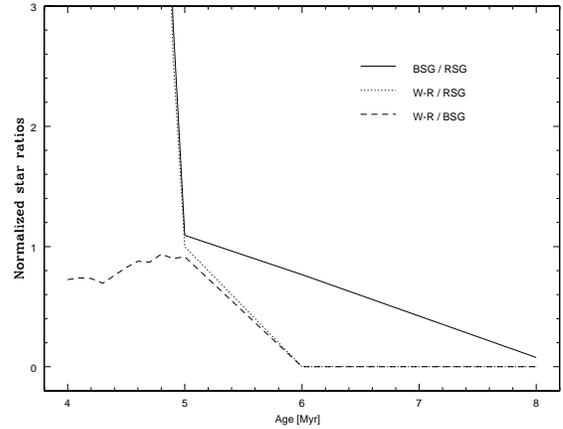}
    \caption{For a cluster model generated with Starburst99, v.5.0  
\citep{VazquezLeitherer05},
    we assumed solar metallicity, an instantaneous burst of a 10$^6$M$_\odot$  
cluster, and a Kroupa IMF.
    For cluster ages between
    4 and 5 Myrs, we determined the expected star ratios from the  
output HR diagram
    in time steps of 0.1 Myrs (after 5 Myrs in steps of 1 Myr). These  
theoretical values were,
    for each time step, divided by the observed ratios determined from
    these absolute star counts: RSG=4, BSG$\geq$25, W-R-stars: 24,
    This means that for each individual diagnostic line, the model  
age is where this
    normalized ratio is 1. For all three diagnostic lines this occurs  
at 5 Myrs.
       \label{agedet}}
\end{figure}

 From stellar number counts, we also (like \citet{Clarketal2005})  
arrive at an expected total
cluster mass of approximately
10$^5$M$_{\odot}$. We see 4 red supergiants, and take
the number of Wolf-Rayet stars (WN: 16, WC: 8) from \citet{Crowtheretal2006}, and of
blue supergiants (BSG; $\geq 25$) from \citet{Clarketal2005}.
The relative numbers of these stellar populations can be used  
together with evolutionary synthesis models
\citep{VazquezLeitherer05} to age date the cluster.
The model we assumed was the same we described earlier in this
section.
 From this, according to Fig. \ref{agedet}, we determine the best  
fitting age to be 5 Myrs.
The absolute stellar numbers then indeed correspond to a 10$^5$M$_{\odot}$ cluster.
Uncertainties on cluster mass from this method are difficult to  
determine, because of
the sudden appearance (or disappearance) of post-main-sequence stars,  
RSG and Wolf-Rayet
stars being almost mutually exclusive. Nevertheless, if the cluster  
age was wrong by
one Myr, the total cluster mass could be as small as 5$\times10^4$M$_{\odot}$.
One puzzling feature remains that supergiant stars of spectral type
B3Ia were reported in \citet{Clarketal2005}, which according to the  
model only appear at
a cluster age of around 12 Myrs. This could indicate that the  
assumption of an
instantaneous burst is invalid.

\section{Conclusions and future work}

    \begin{enumerate}
       \item We identify four late-type supergiant stars (M3I-M6I) which
	in total contribute 40\% of the light in K-band in the scanned region.
	The equivalent width of the CO bandhead of the integrated 
	cluster spectrum is 8.7\AA, which translates to
	a model age of 6.3 Myrs, somewhat older than the 5 Myrs determined from
	the stellar content. This discrepancy could arise from uncertainties in
	the models (the same model is used for both methods of age dating),
         or, certainly in part, because some of the stars which would
	contribute to the diluting continuum are not included in the  
spectrum because
	they are either outside the scanned region, or extinguished by dust.
       \item The velocity dispersion as determined from the
	integrated spectrum is below 7\,km/s; the rms velocity dispersion  
from individual
	stars was found to be 5.8$\pm2.1$\,km/s. The resulting
	dynamical mass is $6.3^{\scriptstyle{+5.3}}_{\scriptstyle{-3.7}}\times 10^4\,M_{\odot} $.
	Of course, these results
	cannot be considered final, because there are at least three aspects  
which could
	influence the velocity dispersion. Firstly, of course the uncertainties
	are rather large for an rms value determined from only four stars,  
and we
	need more stars for a more reliable estimate.
	Secondly, mass segregation is
	not ruled out in this cluster, since the main sequence has not yet been
	detected. In that case, even measuring more stars would yield a
	velocity dispersion which is too low for the total cluster mass,
because the mass segregated high-mass stars have a smaller-than-average
velocity, and also velocity dispersion. Furthermore, these most luminous
cluster members would cause an artificially low estimate of the cluster size.
	In that case we would have a clear demonstration of the perils of  
directly interpreting stellar velocity
	dispersion measurements as reliable mass indicators in more distant  
clusters.
	And thirdly, this cluster is at the lower mass limit
	where observables become relatively unaffected by statistical
	fluctuations in the sampling of the IMF \citep{Cervino2003,Cervino2006},
	which is why we observed it.
	Nevertheless, its probable mass, and hence the number of high mass
	stars, is only 1/10 that of some of the high mass extragalactic star
	clusters.
      \item There is not yet a sign for rapid
	expansion or collapse, which would have shown up as an unexpectedly  
        large velocity dispersion. The assumption of Virial equilibrium thus 
	appears to not be strongly violated.
       \item As this observational technique and analysis works well,  
	 further work at a wavelength more sensitive to the far more numerous 
	 hotter stars in this cluster would greatly increase the
          significance of our preliminary findings.

    \end{enumerate}

\begin{acknowledgements}
We are enormously grateful to Paranal staff, whose excellent support
enabled us to observe in this non-standard technique, and to the
anonymous referee for useful comments which helped us to improve,
and particularly clarify, the paper.
\end{acknowledgements}

\end{document}